\documentclass[aps,prl,showpacs,twocolumn,amsmath,amssymb,superscriptaddress]{revtex4}
\usepackage{graphicx}
\usepackage{bm}
\usepackage{amssymb}
\usepackage{amsmath}
\usepackage{latexsym}
\usepackage{color}
\newcommand{\vc}[1]{\mbox{\boldmath $#1$}} 
\newcommand{\ind}[1]{_{#1}}    
\newcommand{\indrm}[1]{_{\mathrm {#1}}}    

\newcommand{\cddw}{CD\filter DW}   
\newcommand{\filter}{F} 
\newcommand{\pathlength}{{\cal F}}   
\newcommand{\CT}{Czerny-Turner}   
\newcommand{\dirate}{{\mathcal D}}   

\newcommand{\dirateo}{{\mathcal D}_{\indrm{in}}}   
\newcommand{\dirateh}{{\mathcal D}_{\indrm{out}}}   
\newcommand{\dirateall}[1]{\dirate_{\cup_{\ind{#1}}}}
\newcommand{\dirateexp}[1]{\dirate_{\indrm{CDFDW}}}

\begin{document}  
\title{Hard X-ray Spectrographs with Resolution  Beyond $100~\mu$eV}

\author{Yuri Shvyd'ko}
\email{shvydko@aps.anl.gov}
\affiliation{Advanced Photon Source, Argonne National Laboratory,   Argonne, Illinois 60439, USA} 
\author{Stanislav Stoupin}\affiliation{Advanced Photon Source, Argonne National Laboratory,  Argonne, Illinois 60439, USA} 
\author{Kiran Mundboth} \affiliation{ Diamond Light Source Ltd, Didcot  Oxfordshire OX11 0DE, UK}
\author{Jungho Kim} \affiliation{Advanced Photon Source, Argonne  National Laboratory, Argonne, Illinois 60439, USA} 

\begin{abstract} 
  Spectrographs take snapshots of photon spectra with array detectors
  by dispersing photons of different energies into distinct directions
  and spacial locations. Spectrographs require optics with a large
  angular dispersion rate as the key component.  In visible light
  optics diffraction gratings are used for this purpose. In the hard
  x-ray regime, achieving large dispersion rates is a challenge. Here
  we show that multi-crystal, multi-Bragg-reflection arrangements
  feature cumulative angular dispersion rates almost two orders of
  magnitude larger than those attainable with a single Bragg
  reflection. As a result, the multi-crystal arrangements become
  potential dispersing elements of hard x-ray spectrographs.  The hard
  x-ray spectrograph principles are demonstrated by imaging a spectrum
  of photons with a record high resolution of $\Delta E \simeq
  90~\mu$eV in hard x-ray regime, using multi-crystal optics as
  dispersing element. The spectrographs can boost research using
  inelastic ultra-high-resolution x-ray spectroscopies with
  synchrotrons and seeded XFELs.
\end{abstract}

\pacs{42.25.-p, 41.50.+h, 07.85.Nc, 78.70.Ck, 07.85.Fv}


\maketitle

A dream x-ray spectrometer is actually a spectrograph that images
x-ray spectra in one shot, and with an ultimate spectral
resolution. State of the art single shot x-ray spectrometers
\cite{MP81,HVA05,YHZ06,ZCF12} are imaging spectra with array
detectors, using Bragg's law dispersion (BD). BD links the angle of
incidence $\theta$ to the energy $E$ of photons Bragg reflected from
the crystal atomic planes. However, the spectral resolution of the
BD-spectrometers is always limited by the Bragg reflection (Darwin)
bandwidth.

\begin{figure*}[t!]
\setlength{\unitlength}{\textwidth}
\begin{picture}(1,0.30)(0,0)
\put(0.025,0.00){\includegraphics[width=0.95\textwidth]{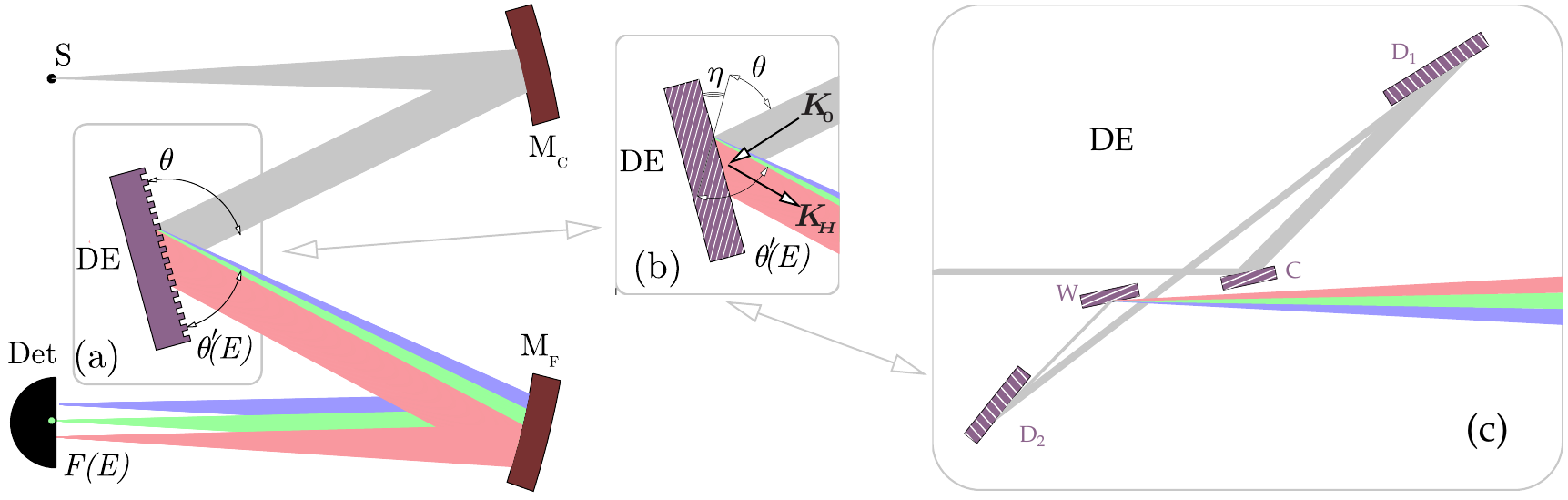}}
\end{picture}
\caption{Scheme of the \CT\ type \cite{CT30} spectrograph with a
  diffraction grating (a), or a crystal in asymmetric x-ray Bragg
  diffraction (b) as dispersing element - DE. Other components include
  radiation source S, collimating and focusing mirrors M$_{\indrm{C}}$
  and M$_{\indrm{F}}$, and position sensitive detector Det. (c)
  Multi-crystal multi-reflection \cddw\ optics - is an example of a
  hard x-ray ``diffraction grating'' (DE element) with enhanced
  dispersion rate, suitable for hard x-ray spectrographs.}
\label{fig001}
\end{figure*}

Angular dispersion (AD) is one way how to overcome the Darwin width
limitation and substantially improved spectral resolution of x-ray
optics \cite{Shvydko-SB,SLK06}.  AD is a variation of the photon angle
of reflection $\theta^{\prime}$, for a fixed incidence angle $\theta$,
with the photon energy $E$.  AD takes place in Bragg diffraction,
albeit only if the diffracting atomic planes are at a nonzero angle
$\eta\not =0$ (asymmetry angle) to the entrance crystal surface
\cite{MK80-1,BSS95,Shvydko-SB}, see Fig.~\ref{fig001}(b).

Unlike BD, AD links $\theta^{\prime}$ to $E$ for a fixed $\theta$. AD
is independent of the Darwin width, and can be therefore used to
resolve much narrower spectral features. Using angular-dispersive
monochromators,
x-rays were already monochromatized to bandwidths (0.45~meV) almost
two orders of magnitude smaller than the width of the Bragg
reflections (27~meV) involved \cite{ShSS11}.
New concepts are required, however, to realize single shot
angular-dispersive spectrographs.

We show here that multi-Bragg-reflection arrangements feature, in
theory and in experiment, cumulative angular dispersion rates almost
two orders of magnitude greater than those attainable in a single
Bragg reflection. An angular-dispersive x-ray spectrograph of a \CT
-type \cite{CT30} is introduced with the enhanced angular-dispersive
optics as a ``diffraction grating''. A record high spectral resolution
of $\Delta E \simeq 90~\mu$rad is demonstrated in the hard x-ray regime.

\CT\ grating spectrographs are nowadays standard in infrared, visible,
and ultraviolet spectroscopies \cite{SMD64, LTR10}. In its classical
arrangement, the spectrographs comprise, first, a collimating mirror
M$_{\indrm{C}}$, which collects photons from a radiation source S and
collimates the photon beam - see Fig.~\ref{fig001}(a); second, an
angular-dispersive element DE such as a diffraction grating or a
prism, which disperses photons of different energies into different
directions; third, a curved mirror M$_{\indrm{F}}$ which focuses
photons of different energies onto different locations $x(E)$, and,
last but not least, a spatially sensitive detector Det placed in the
focal plane to record the whole photon spectrum.  To achieve high
resolution, the most important factor is the magnitude of the AD rate
$\dirate=\delta\theta^{\prime}/\delta E$, which measures the variation
of the reflection angle $\theta^{\prime}$ with photon energy $E$ upon
reflection from the dispersing element.  For the given mirror focal
length $\pathlength$ ($ {\mathrm M}_{\indrm{F}}\rightarrow {\mathrm
  {Det}}$), the AD rate $\dirate$ determines the variation of the
source image position $x(E)$ on the detector with respect to photon
energy: $\delta x(E)=\dirate\pathlength\, \delta E$.
The smallest spectral interval $\Delta E$ which can be resolved is
therefore
\begin{equation}
\Delta E\,=\, \frac{1}{\dirate}\,  \frac{\Delta x}{\pathlength },
\label{eq003}
\end{equation}
where $\Delta x$ is the largest of either the  source S image size on the
detector for a particular monochromatic component or detector spatial
resolution.

Nowadays, diffraction grating manufacturing technology has advanced to
the extent that grating spectrographs are being successfully used
with much shorter wavelengths, in particular in soft x-ray regime
($\lesssim 1$~keV) \cite{GPD06} attaining a resolving power of $E/\Delta
E \simeq 10^4$. Extension into the hard x-ray regime is, however, not
trivial, because of the lack of hard x-ray optics elements with
sufficiently large dispersion rate.

A hard x-ray equivalent of the diffraction grating is a Bragg
diffracting crystal with diffracting atomic planes at an asymmetry
angle $\eta\not =0$ to the entrance crystal surface -
Fig.~\ref{fig001}(b) \cite{MK80-1,BSS95,Shvydko-SB}.  The AD rates in
Bragg diffraction are typically small $\dirate \simeq 8~\mu$rad/meV
\cite{SLK06}, being the main obstacle in realization of hard x-ray
spectrographs \footnote{The use of asymmetrically cut crystal as
  dispersing element in combination with a focusing element was
  proposed for a ``focusing monochromator"\cite{KCR09}. As was noticed
  in \cite{KCR09} such optics can be used as a monochromator, however,
  not as a spectral analyzer (or spectrograph), as a small angular
  size of the radiation source is required for its realization.}. The
AD rate can be enhanced dramatically, by successive asymmetric Bragg
reflections, as shown below.

Let $\vc{K}_{\ind{0}}$ be the momentum of the incident x-ray photon,
and $\vc{K}_{\ind{H}}$ of the photon reflected from a crystal in Bragg
reflection with the diffraction vector $\vc{H}$. The vectors
$\vc{K}_{\ind{0}}$ and $\vc{K}_{\ind{H}}$ make angles $\theta+\eta$,
and $\theta^{\prime}-\eta$, respectively, with the crystal surface -
Fig.~\ref{fig001}(b). 
The asymmetry angle $\eta$ is defined here to be positive in the
geometry shown in Fig.~\ref{fig001}(b), and negative in the geometry
with reversed incident and reflected x-rays (not shown).  Conservation
of the tangential components
$(\vc{K}_{\ind{H}})_t=(\vc{K}_{\ind{0}}+\vc{H})_t$ with respect to the
entrance crystal surface, and the conservation of the photon energies
$|\vc{K}_{\ind{H}}|\hbar c=|\vc{K}_{\ind{0}}|\hbar c=K\hbar c= E$
require that \cite{KB76}:
\begin{equation}                                                     
  \cos(\theta^{\prime}-\eta) =  \cos(\theta+\eta) + \frac{H}{K}\sin{\eta}.                
\label{ad000}                                                        
\end{equation}                                                       
Differentiating over $E$, using Bragg's law $2K\sin\theta=H$, and
assuming $|\theta^{\prime}-\theta|\ll 1$, we obtain
\begin{equation}                                                                                                                    
  \frac{{\mathrm d}\theta^{\prime}}{{\mathrm d}E} = -b \frac{{\mathrm d}\theta}{{\mathrm d}E} +\dirate, \hspace{0.5cm}\dirate=\frac{2\sin\theta \sin{\eta}}{E\sin(\theta-\eta)}         
\label{ad010}
\end{equation}
Here $b=-\sin(\theta+\eta)/\sin(\theta-\eta)$ is the asymmetry
ratio. If the incident beam is collimated, ${{\mathrm
    d}\theta}/{{\mathrm d}E}=0$, then ${{\mathrm
    d}\theta^{\prime}}/{{\mathrm d}E}=\dirate$, where $\dirate$
\eqref{ad010} is the intrinsic AD rate in a Bragg
reflection \cite{MK80-1,Shvydko-SB}. If, however, the incident x-rays
are dispersed, ${{\mathrm d}\theta}/{{\mathrm d}E}\not=0$, then the
dispersion rate ${{\mathrm d}\theta^{\prime}}/{{\mathrm d}E}=\dirateh$ becomes
\begin{equation}                                                                                                                    
\dirateh = b \dirateo +\dirate ,         
\label{ad020}
\end{equation}
where $\dirateo = - {{\mathrm d}\theta}/{{\mathrm d}E}$.  The minus
sign follows the convention that the counterclockwise sense of angular
variations in $\theta$ and $\theta^{\prime}$ is positive.  Similarly,
if the sense of deflection of the ray upon the Bragg reflection is
clockwise, unlike that shown in Fig.~\ref{fig001}(b), then $\dirate$
in Eq.~\eqref{ad010} has to be used with sign minus.

Equation~\eqref{ad020} demonstrates that the AD rate $\dirateo$ can
be indeed significantly enhanced by two successive asymmetric Bragg
reflections, if its asymmetry ratio is large: $|b|\gg 1$. The
enhancement can be even larger if several ($1,2, .., n$) successive
reflections are used:
\begin{multline}                                                                                                                    
  \dirateall{n}\,=\, b_{\ind{n}}\dirateall{n-1} + \dirate_{\ind{n}}\,= \\ b_{\ind{n}}
  (b_{\ind{n-1}} \dotso (b_{\ind{3}}(b_{\ind{2}}\dirate_{\ind{1}} + \dirate_{\ind{2}}) + \dirate_{\ind{3}})\dotso
  \dirate_{\ind{n-1}}) + \dirate_{\ind{n}}.
\label{ad030}
\end{multline}                                                                                                            

\begin{figure*}
\setlength{\unitlength}{\textwidth}
\begin{picture}(1,0.57)(0,0)
\put(0.01,0.0){\includegraphics[width=0.96\textwidth]{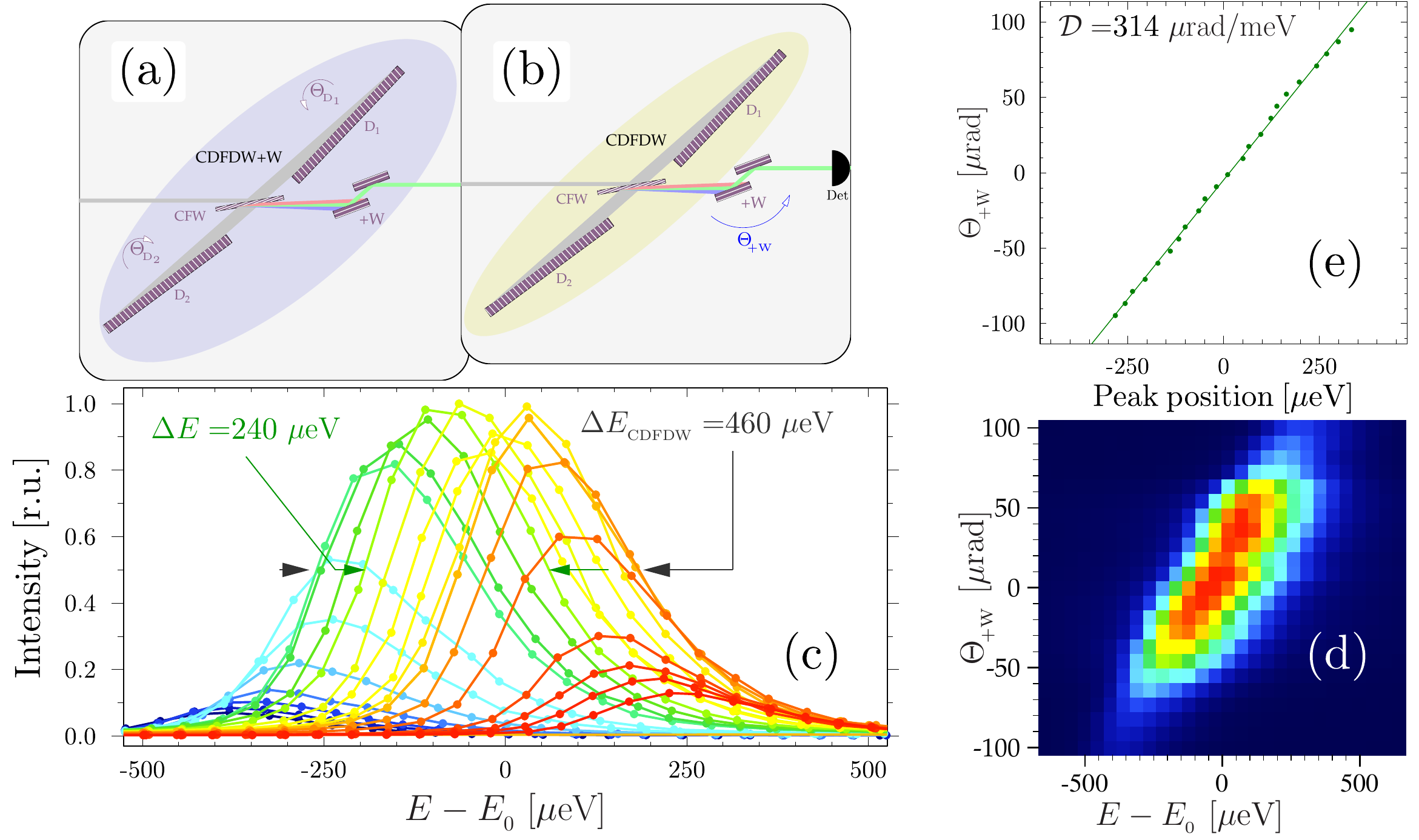}}
\end{picture}
\caption{Enhanced angular dispersion rate of the \cddw\ optics. 
  Schematic of the experiment showing the CDFDW+W monochromator (a)
  probing the angular dispersion rate of the CDFDW optics (b) with
  the +W channel-cut crystal as an angular analyzer. (c) Spectral
  dependences of x-ray transmission through the CDFDW optics for fixed
  angular positions $\Theta_{\indrm{+W}}$ of +W. (d) Same as (c)
  presented as a 2D plot. (e) Transmission peak position as a function
  of $\Theta_{\indrm{+W}}$.}
\label{fig002}
\end{figure*}

In the first experiment presented below, we demonstrate this effect on
an example of a four-crystal angular dispersive CDFDW optics, with
schematic shown in Fig.~\ref{fig001}(c).  In the second proof of
principle experiment presented in this paper, we apply such optics as
the ``dispersion grating'' of a prototype hard x-ray spectrograph to
image a spectrum of the CDFDW with record high spectral resolution.
The experiments were performed at 30ID beamline of the APS.

Details on the CDFDW optics used in this paper are provided in
\cite{ShSS11}. CDFDW is a modification of the CDW
\cite{Shvydko-SB,SLK06} optics originally designed to achieve the very
high monochromatization of x-rays.  The first element - C (collimator)
- is a Si asymmetrically cut crystal, with the 220 Bragg reflection,
$\theta_{\indrm{C}}=20.7^{\circ}$, $\eta_{\indrm{C}}=19.0^{\circ}$,
$b_{\indrm{C}}=-1/21.5$, accepting x-rays with photon energy
$E=9.1315$~keV in a wide angular range $\simeq 110~\mu$rad, and
collimating it to a beam with a $|b_{\indrm{C}}|$ smaller divergence,
and negligible $\dirate_{\indrm{C}}=0.040~\mu$rad/meV. The next two Si
crystals - D$_{\ind{1}}$ and D$_{\ind{2}}$, are designed to produce
maximal intrinsic Bragg dispersion rate $\dirate$ \eqref{ad010}, using
the 008 reflection with $\theta_{\indrm{D_i}}\simeq 90^{\circ}$,
$\eta_{\indrm{C_i}}=88.0^{\circ}$, $b_{\indrm{D_i}}\simeq -1$
($i=1,2$), and
$\dirate_{\indrm{D_1}}=-\dirate_{\indrm{D_2}}=6.27~\mu$rad/meV. Note,
that the scheme in Fig.~\ref{fig001}(c) is shown in a generic
configuration with $\theta_{\indrm{D_i}}\not = 90^{\circ}$. The fourth
crystal - W, is equivalent to the C-crystal, however applied in an
inverse configuration with $\eta_{\indrm{W}}=-\eta_{\indrm{C}}$,
$b_{\indrm{W}}=1/b_{\indrm{C}}=-21.5$, and
$\dirate_{\indrm{W}}=0.86$. It is used to enhance the AD rate of the
D-crystals. Indeed, applying Eq.~\eqref{ad030} we estimate the
cumulative dispersion rate of the optics: $\dirateall{\mathrm
  {}}\simeq b_{\indrm{W}}(\dirate_{\indrm{D_2}}
-\dirate_{\indrm{D_1}})+\dirate_{\indrm{W}} \simeq
2b_{\indrm{W}}\dirate_{\indrm{D_1}} \simeq -270~\mu$rad/meV, enhanced
by a factor $2|b_{\indrm{W}}|\simeq 43$ compared to the dispersion
rate achieved in a single Bragg reflection. The same enhancement
factor was derived in \cite{Shv11} using DuMond diagram analysis.

Figure~\ref{fig002} shows a schematic of the first experiment and
results of measurements of the AD rate of the \cddw\ optics.  A
tunable monochromator with a $\simeq 170~\mu$eV bandwidth shown
schematically in Fig.~\ref{fig002}(a), and described in detail in the
next paragraph, is used to measure transmission spectra through the
\cddw\ optics under study, presented in Fig.~\ref{fig002}(b).  An
auxiliary element denoted as +W in Fig.~\ref{fig002}(b), is used to
extract from all x-ray photons emanating from the \cddw\ optics a
small part with $\simeq 20~\mu$rad divergence by the 220 symmetric
Bragg reflection (angular acceptance $20~\mu$rad) from a Si
channel-cut crystal. For each angular position $\Theta_{\indrm{+W}}$
of the +W crystal, a spectrum of x-rays transmitted through the \cddw\
optics and through the +W angular analyzer is measured, as shown in
Fig.~\ref{fig002}(c).  Figure~\ref{fig002}(d) presents a 2D plot of
the spectra.  The peak of the spectral distribution changes with the
emission angle defined by $\Theta_{\indrm{+W}}$.
Figure~\ref{fig002}(e) shows that the dependence is linear, with the
tangent $\dirateexp\ = 314~\mu$rad/meV representing the measured
dispersion rate of the \cddw\ optics.  The number is even higher than
the previously estimated one, which we attribute to the difference
between the nominal and real asymmetry angles. The result confirms the
theoretical prediction, expressed by Eqs.~\eqref{ad020}-\eqref{ad030},
that the angular dispersion rate can be substantially enhanced in
multi-crystal arrangements.

A few details regarding the monochromator in Fig.~\ref{fig002}(a) are
in order: It is the same CDFDW optics, however, enhanced with the +W
channel-cut that substantially decrease the CDFDW bandwidth.

The energy tuning of the CDFDW+W monochromator is performed by
synchronous change of the angular orientation of the D-crystals, as
indicated by $\Theta_{\indrm{D_i}}$ in Fig.~\ref{fig002}(a), and
explained in more detail in \cite{ShSS11}.  Each spectral dependence
in Figure~\ref{fig002}(c) has a width of $\Delta E\simeq 240~\mu$eV.
This number represents the width of the convolution of the spectral
distributions of the CDFDW+W monochromator - Fig.~\ref{fig002}(a), and
the analyzer - Fig.~\ref{fig002}(b). Assuming they are equivalent, the
spectral width of a single CDFDW+W optics is estimated as $\Delta
E/\sqrt{2}\simeq 170~\mu$eV. An envelope of spectral dependences in
Fig.~\ref{fig002}(c) reveals a total width of $\Delta
E_{\indrm{\cddw}}\simeq 460~\mu$eV, and represents the spectral width
of the CDFDW optics. A similar number has been measured in
\cite{ShSS11}.

\begin{figure}
\setlength{\unitlength}{\textwidth}
\begin{picture}(1,0.44)(0,0)
  \put(0.0,0.0){\includegraphics[width=0.50\textwidth]{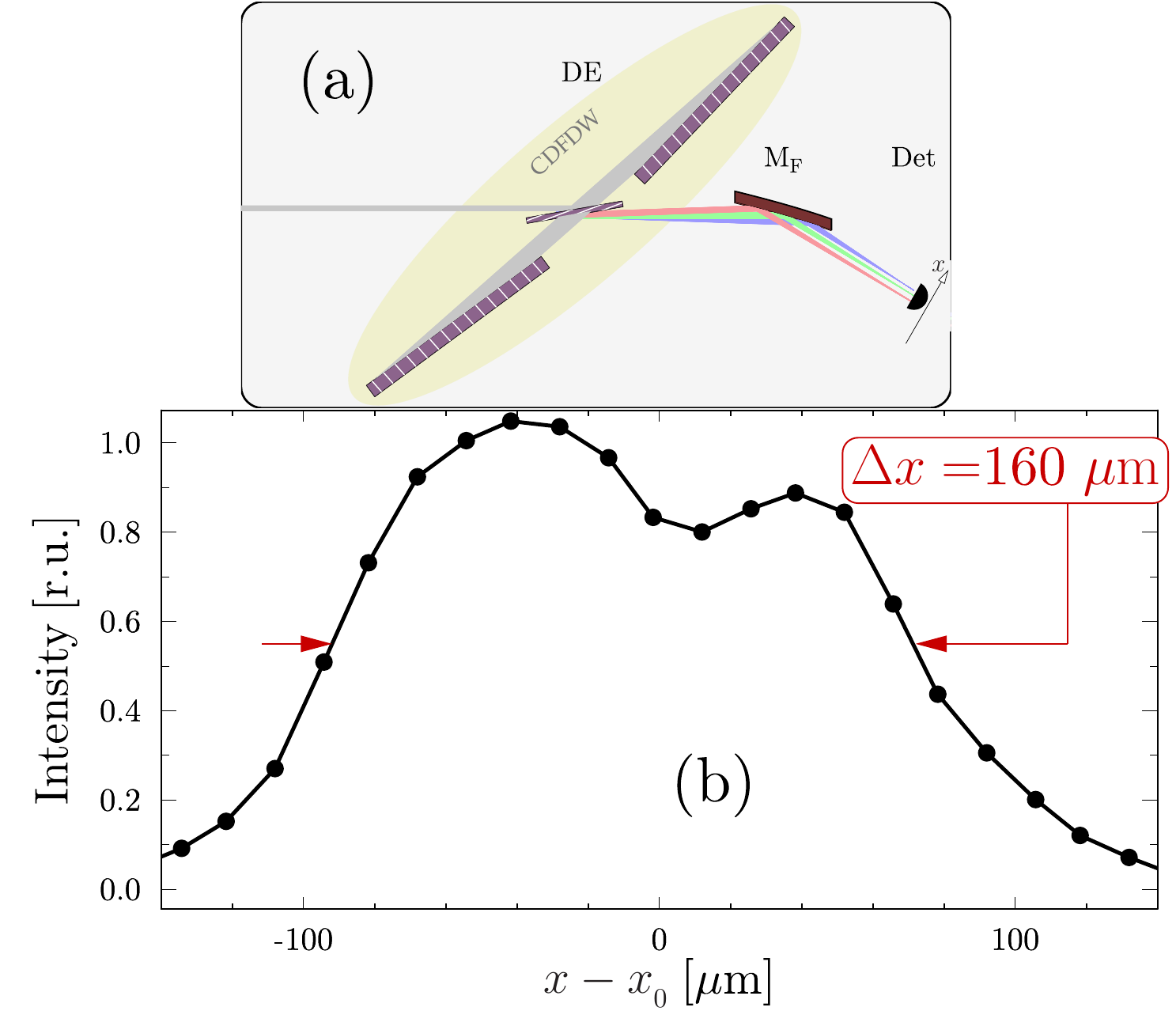}}
\end{picture}
\caption{Image of the CDFDW spectral function on the spatial $x$-scale
  (b) measured in the prototype spectrograph  setup  (a). }
\label{fig003}
\end{figure}

\begin{figure*}
\setlength{\unitlength}{\textwidth}
\begin{picture}(1,0.54)(0,0)
  \put(0.02,0.00){\includegraphics[width=0.90\textwidth]{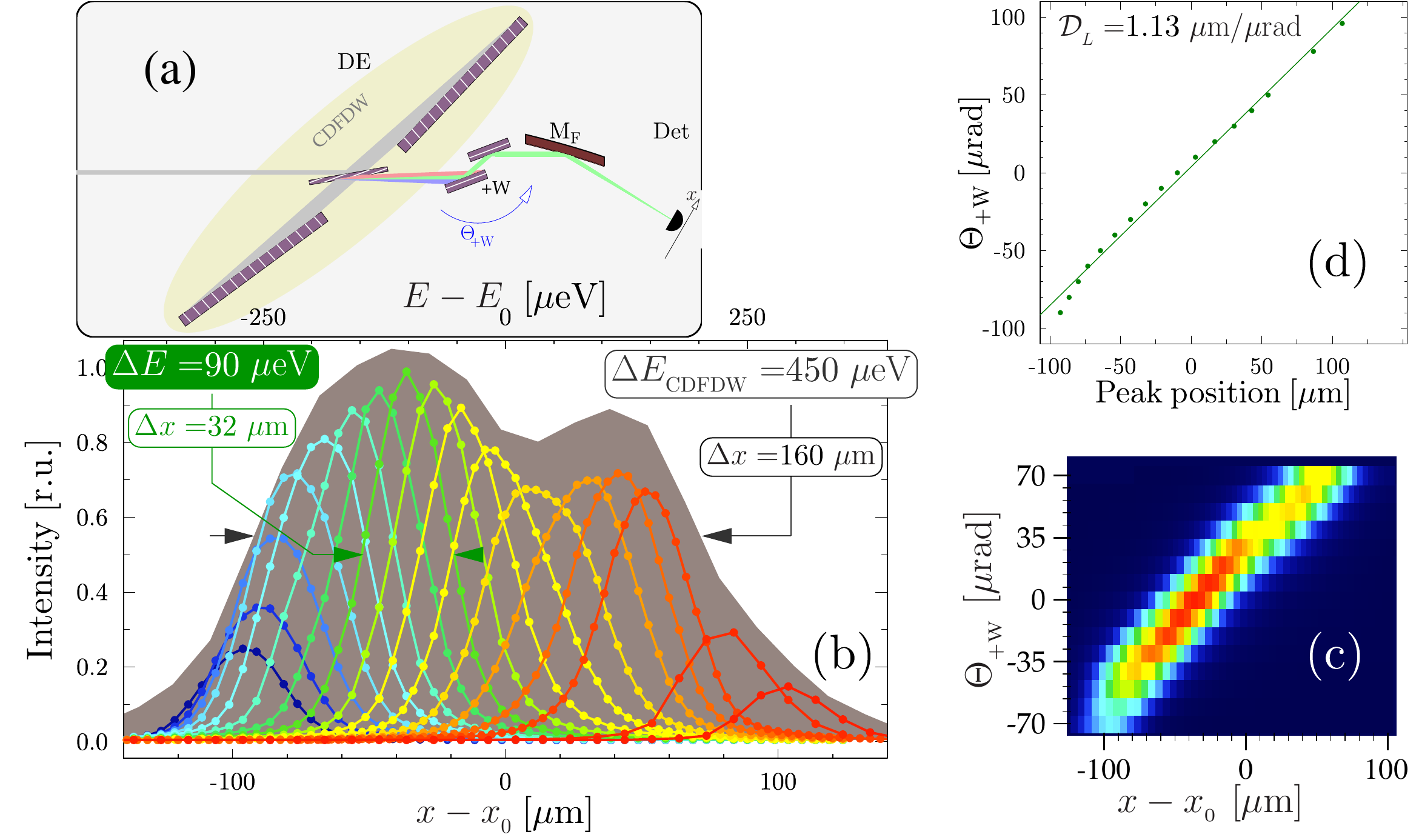}}
\end{picture}
\caption{Prototype hard x-ray spectrograph with a $\simeq 90\mu$eV
  resolution using CDFDW optics as dispersing element (DE).  (a) Schematic
  of the experiment. (b) Images of the CDFDW+W spectral function on
  the spatial $x$-scale for fixed angular positions
  $\Theta_{\indrm{+W}}$ of +W. (c) Same as (b) presented as a 2D
  plot. (d) Spatial peak position as a function of
  $\Theta_{\indrm{+W}}$.}
\label{fig004}
\end{figure*}

In the second experiment, the CDFDW optics with the demonstrated above
greatly enhanced dispersion rate is used as a ``dispersion grating''
of a prototype hard x-ray spectrograph.  The enhanced dispersion rate
enables imaging of the CDFDW spectrum with a record high spectral
resolution.  Figure~\ref{fig003}(a) shows the proof of the principle
spectrograph setup, in which the CDFDW optics is augmented with a
focusing mirror ${\mathrm M}_{\mathrm F}$ and a position sensitive
detector placed at the mirror focal distance ${\mathcal
  F}=1.38$~m. The polychromatic collimated beam, which in a complete
spectrograph setup would be created by the collimating mirror
${\mathrm M}_{\mathrm C}$, as in Fig.~\ref{fig001}(a), is mimicked
here by the incident x-ray beam with a $\lesssim 15~\mu$rad angular
divergence and a bandwidth of $\simeq 0.6$~eV. This beam is focused by
mirror ${\mathrm M}_{\mathrm F}$ to produce a spot with a $\simeq
30~\mu$m size.  Figure~\ref{fig003}(b) shows the spatial distribution
of x-rays transmitted through the CDFDW optics and imaged on the
detector with mirror ${\mathrm M}_{\mathrm F}$. It is $\simeq
160~\mu$m broad, i.e., much broader than the $\simeq 30~\mu$m image
size of the incident beam, and has a double peak structure.  We
believe that this distribution presents the image of the spectrum of
x-rays transmitted through the CDFDW optics, mapped on $x$ using
mirror ${\mathrm M}_{\mathrm F}$. It correspond to the envelope of
spectral dependences shown in Fig.~\ref{fig002}(c), however with a
more pronounce minimum near $x=0$, as no convolution with the
monochromator spectral function is involved.  The spatial width
$\Delta x \simeq 160~\mu$m corresponds to the spectral width, which we
know should be $\simeq 450~\mu$eV.  The results shown in
Fig.~\ref{fig004}(b), which were measured in a modified experimental
scheme presented in Fig.~\ref{fig004}(a), confirm this.

The experiential scheme in Fig.~\ref{fig004}(a) is complemented by the
+W channel-cut which, as we know from the results presented in
Fig.~\ref{fig002}, selects x-ray emanating in a certain direction from
the CDFDW optics, and, as a result, within a reduced bandwidth, whose
central photon energy is defined by the angle $\Theta_{\indrm{+W}}$.
Figure~\ref{fig004}(b) shows the spatial distribution of x-rays
measured at different $\Theta_{\indrm{+W}}$ values. The peak
positions, plotted in Figure~\ref{fig004}(d), change linearly with
$\Theta_{\indrm{+W}}$ at a rate ${\mathcal
  D_{\ind{F}}}=1.13~\mu$m/$\mu$rad.  Together with the results
presented in Fig.~\ref{fig002}(e) this proves that the spectral
distribution of x-rays from the CDFDW optics is imaged by mirror
${\mathrm M}_{\mathrm F}$ on the spatial scale, with a conversion
factor ${\mathcal D_{\ind{F}}}\dirateexp\ \simeq 355~\mu$m/meV. Using
this number, we obtain that the total widths of the spatial
distributions in Figs.~\ref{fig003}(b) and \ref{fig004}(b) are
$\Delta_{\indrm{CDFDW}}\simeq 450~\mu$eV, representing the spectral
width of the CDFDW optics. The spatial widths of single lines vary
from $32~\mu$m (in green) to $50~\mu$m (in yellow), corresponding to
spectral widths $\simeq 90~\mu$eV and $\simeq 140~\mu$eV
respectively. The resolution of the CDFDW spectrograph is at least
$\simeq 90~\mu$eV, or better. The fact that this value changes across
the CDFDW spectrum, as well as the fact that the CDFDW spectrum has a
double-peak structure, imply that the CDFDW optics we have built is
not yet perfect. This is consistent with the results of \cite{ShSS11},
where a somewhat broader line was measured as expected from
theory. However, now using the CDFDW in the spectrograph setup, we can
measure and analyze the CDFDW spectrum directly, without the need of 
another CDFDW optics as an analyzer. Using Eq.~\eqref{eq003} with
${\mathcal F}=1.38$~m, $\Delta x = 32~\mu$m, and the theoretically
estimated $\dirateall{}=270~\mu$rad/meV, we obtain $\Delta E
=86~\mu$eV in agreement with the measured energy resolution.

In conclusion, a principle is proposed and demonstrated how to enhance
by more than an order of magnitude the angular dispersion rate of
x-rays in Bragg diffraction, namely by successive asymmetric Bragg
reflections. This effect opens an opportunity of realizing dispersing
elements in the hard x-ray regime with an angular dispersion rate
sufficiently large for x-ray spectrographs. The hard x-ray
spectrograph principle is demonstrated with the multi-crystal
multi-reflection CDFDW optics as dispersing element, by imaging an
x-ray spectrum of 9.1315~keV photons in a $450~\mu$eV window with a
record small $90~\mu$eV resolution, thus achieving spectral resolution
power beyond $10^8$ in hard x-ray regime. The main future effort
should be directed not only to further improving the spectral
resolution, but primarily into making the spectral window broader, to
enhance the spectrographs' throughput. Hard x-ray spectrographs can
advance significantly research using high-resolution x-ray
spectroscopies, in particular different branches of inelastic x-ray
scattering, using synchrotrons and seeded XFELs \cite{HXRSS12}.

  We are grateful to L.~Young for supporting this project at the APS,
  to S.~Collins and G.~Materlik at the DLS. D.~Shu, T.~Roberts,
  K.~Goetze, J.~Kirchman, P.~Jemian, M.~Upton, and Y.~Ding are
  acknowledged for technical support. R.~Lindberg and X.~Yang are
  acknowledged for reading the manuscript and valuable suggestions.
  Work was supported by the U.S. Department of Energy, Office of
  Science, Office of Basic Energy Sciences, under Contract
  No. DE-AC02-06CH11357.

\end{document}